\documentstyle[aps,preprint,psfig]{revtex}

\begin{document}
\title{Error and attack tolerance of complex networks}
\author{R\'eka Albert, Hawoong Jeong, Albert-L\'aszl\'o Barab\'asi}
\address{Department of Physics, University of Notre Dame, Notre Dame, IN 46556}
\maketitle 
 
\vspace{0.5in}

{\bf Many complex systems display a surprising degree of
 tolerance against errors. For example, relatively simple organisms grow, persist and reproduce despite drastic pharmaceutical or environmental interventions, an error tolerance attributed to the robustness of the underlying metabolic network\cite{biology}. Complex  communication networks\cite{claffy} display a surprising degree of robustness: while key components regularly malfunction, local failures rarely lead to the loss of the global information-carrying ability of the network. The stability of these and other complex systems is often attributed to the redundant wiring of the functional web defined by the systems' components. In this paper we demonstrate that error tolerance is not shared by all redundant systems, but it is displayed only by a class of inhomogeneously wired networks, called scale-free networks. We find that scale-free networks, describing  a number of systems, such as the World Wide Web (www)\cite{diam,ibm,hubermann}, Internet\cite{faloutsos}, social networks\cite{social} or a cell\cite{jeong}, display an unexpected degree of robustness, the ability of their nodes to communicate being unaffected by even unrealistically high failure rates. However, error tolerance comes at a high price: these networks are extremely vulnerable to attacks, i.e. to the selection and removal of a few nodes that play the most important role in assuring the network's connectivity. Such error tolerance and attack vulnerability are generic properties of communication networks,  such as the Internet or the www, with complex implications on assuring information readiness.}
 
\newpage

The increasing availability of topological data on large networks, aided by the computerization of data acquisition, has lead to major advances in our understanding of the generic aspects of network structure and development\cite{random,random1,small_world,comp,eco,banavar,banavar1,amaral}. The  existing empirical and theoretical results indicate that complex networks can be divided into two major classes based on their connectivity distribution $P(k)$, giving the probability that a node in the network is connected to $k$ other nodes. The first class of networks is characterized by a $P(k)$ that is peaked at an average $\langle k\rangle$ and decays exponentially for large $k$. The most investigated examples of such {\it exponential networks} are the random graph model of Erd\H{o}s and R\'enyi\cite{random,random1} and the small-world model of Watts and Strogatz\cite{small_world}, both leading to a fairly homogeneous network, in which each node has approximately the same number of links, $k\simeq\langle k\rangle$. In contrast, results on the world-wide web (www)\cite{diam,ibm,hubermann}, Internet\cite{faloutsos} and other large networks\cite{ba,physica,redner} indicate that many systems belong to a class of inhomogeneous networks, referred to as {\it scale-free networks}, for which $P(k)$ decays as a power-law, i.e. $P(k)\sim k^{-\gamma}$, free of a characteristic scale. While the probability that a node has a very large number of connections ($k>>\langle k \rangle$) is practically prohibited in exponential networks, highly connected nodes are statistically significant in scale-free networks (see Fig.$\,$1).

We start by investigating the robustness of the two basic network models, the Erd\H{o}s-R\'enyi (ER) model\cite{random,random1} that produces a network with an exponential tail, and the scale-free model\cite{ba} with a power-law tail. In the ER model we first define the $N$ nodes, and then connect each pair of nodes with probability $p$. This algorithm generates a homogeneous network (Fig.$\,$1), whose connectivity follows a Poisson distribution  peaked at $\langle k\rangle$ and decaying exponentially for $k>>\langle k\rangle$. 

The inhomogeneous connectivity distribution of many real networks is reproduced by the scale-free model\cite{ba,physica} that incorporates two ingredients common to real networks: growth and preferential attachment. The model starts with $m_0$ nodes. At every timestep $t$ a new node is introduced, which is connected to $m$ of the the already existing nodes. The probability $\Pi_i$ that the new node is connected to node $i$ depends on the connectivity $k_i$ of that node, such that $\Pi_i=k_i/ \sum_j k_j$. For large $t$ the connectivity distribution is a power-law following $P(k)=2m^2/k^3$.

The interconnectedness of a network is described by its diameter $d$, defined as the average length of the shortest paths between any two nodes in the network. The diameter characterizes the ability  of two nodes to communicate with each other: the smaller $d$ is, the shorter is the expected path between them. Networks with a very large number of nodes can have a rather small diameter; for example the diameter of the www, with over $800$ million nodes\cite{giles}, is around $19$\cite{diam}, while social networks with over six billion individuals are believed to have a diameter of around six\cite{milgram}. To properly compare the two network models we generated networks that have the same number of nodes and links such that $P(k)$ follows a Poisson distribution for the exponential, and a power-law for the scale-free network.
 
{\it Error tolerance---} To address the networks' error tolerance, we study the changes in the diameter when a small fraction $f$ of the nodes is removed. The malfunctioning (absence) of a node  in general increases the distance between the remaining nodes, since it can eliminate some paths that contribute to the system's interconnectedness. Indeed, for the exponential network the diameter increases monotonically with $f$ (Fig.$\,$2a), thus, despite its redundant wiring (Fig.$\,$1), it is increasingly difficult for the remaining nodes to communicate with each other. This behavior is rooted in the homogeneity of the network: since all nodes have approximately the same number of links, they all contribute equally to the network's diameter, thus the removal of each node causes the same amount of damage. In contrast, we observe a drastically different and surprising behavior for the scale-free network (Fig.$\,$2a): {\it the diameter remains unchanged under an increasing level of errors}. Thus even when as high as $5\%$ of the nodes fail, the communication between the remaining nodes in the network is unaffected. This robustness of scale-free networks is rooted in their extremely inhomogeneous connectivity distribution: since the power-law distribution implies that the majority of nodes have only a few links, nodes with small connectivity will be selected with much higher probability, and the removal of these "small" nodes  does not alter the path structure of the remaining nodes, thus has no impact on the overall network topology. 

{\it Attack survivability---} An informed agent that attempts to deliberately damage a network, such as designing a drug to kill a bacterium, will not eliminate the nodes randomly, but will rather target the most connected nodes. To simulate an attack we first remove the most connected node, and continue selecting and removing nodes in the decreasing order of their connectivity $k$. Measuring the diameter of an exponential network under attack, we find that, due to the homogeneity of the network, there is no substantial difference whether the nodes are selected randomly or in decreasing order of connectivity (Fig.$\,$2a). On the other hand, a drastically different behavior is observed for scale-free networks: when the most connected nodes are eliminated, the diameter of the scale-free network increases rapidly, doubling its original value if $5\%$ of the nodes are removed. This vulnerability to attacks is rooted in the inhomogeneity of the connectivity distribution: the connectivity is ensured by a few highly connected nodes (Fig.$\,$1b), whose removal drastically alters the network's topology, and decreases the ability of the remaining nodes to communicate with each other.

{\it Network fragmentation---} When nodes are removed from a network, clusters of nodes, whose links to the system disappear, can get cut off from the main cluster. To better understand the impact of failures and attacks on the network structure, we next investigate this fragmentation process. We measure the size of the largest cluster, $S$, shown as a fraction of the total system size, when a fraction $f$ of the nodes are removed either randomly or in an attack mode. We find that for the exponential network, as we increase $f$, $S$ displays a threshold-like behavior such that for $f>f_c\simeq 0.28$ we have $S\simeq 0$. A similar behavior is observed when we monitor the average size $\langle s\rangle$ of the isolated clusters (i.e. all the clusters except the largest one), finding that $\langle s\rangle$ increases rapidly until $\langle s\rangle\simeq 2$ at $f_c$, after which it decreases to $\langle s\rangle=1$. These results indicate the following breakdown scenario (Fig.$\,$4): For small $f$, only single nodes break apart, $\langle s\rangle\simeq 1$, but as $f$ increases, the size of the fragments that fall off the main cluster increases,  displaying a singular behavior at $f_c$. At $f_c$ the system practically falls apart, the main cluster breaking into small pieces, leading to $S\simeq 0$, and the size of the fragments, $\langle s\rangle$, peaks. As we continue to remove nodes ($f>f_c$), we fragment these isolated clusters, leading to a decreasing $\langle s \rangle$. Since the ER model is equivalent to the infinite dimensional percolation\cite{perc}, the observed threshold behavior is qualitatively similar to the percolation critical point.

However, the response of a scale-free network to attacks and failures is rather different (Fig.$\,$3b). For random failures no threshold for fragmentation is observed, rather the size of the largest cluster slowly decreases. The fact that $\langle s\rangle\simeq 1$ for most $f$ indicates that the network is deflated by nodes breaking off one by one, the increasing error level leading to the {\it isolation of single nodes only, not clusters of nodes}. Thus, in contrast with the catastrophic fragmentation of the exponential network at $f_c$, the scale-free network stays together as a large cluster for very high values of $f$, providing additional evidence of the topological stability of these networks under random failures. This behavior is consistent with the existence of an extremely delayed critical point (Fig.$\,$3), the network falling apart
only after the main cluster has been completely deflated. On the other hand, the response to attack of the scale-free network is similar (but swifter) to the response to attack and failure of the exponential network (Fig.$\,$3b): at a critical threshold $f_c^{sf}\simeq 0.18$, smaller than the value $f_c^e\simeq 0.28$ observed for the exponential network, the system breaks apart, forming many isolated clusters (Fig.$\,$4). 

While great efforts are being made to design error tolerant and low yield components for communication systems, little is known about the effect of the errors and attacks on the large-scale connectivity of the network. To demonstrate the impact of our model based studies to these systems, next we investigate the error and attack tolerance of two networks of increasing economic and strategic importance: the Internet and the www.

  Recently Faloutsos {\it et al.}\cite{faloutsos} investigated the topological properties of the Internet at the router and inter-domain level, finding that the connectivity distribution follows a power-law, $P(k)\sim k^{-2.48}$. Consequently, we expect that it should display the error tolerance and attack vulnerability predicted by our study. To test this, we used  the latest survey of the Internet  topology, giving the network at the inter-domain (autonomous system) level. Indeed, we find that the diameter of the Internet is unaffected by the random removal of as high as $2.5\%$ of the nodes (an order of magnitude larger than the failure rate ($0.33\%$) of the Internet routers\cite{labovitz}), while if the same percentage of the most connected nodes are eliminated (attack), $d$ more than triples (Fig.$\,$2b). Similarly, the large connected cluster persists for high rates of random node removal, but if nodes are removed in the attack mode, the size of the fragments that break off increases rapidly, the critical point appearing at $f_c^I\simeq 0.03$ (Fig.$\,$3b). 

The www forms a huge directed graph whose nodes are documents and edges are the URL hyperlinks that point from one document to another, its topology determining the search engines' ability to locate information on it. The www is also a scale-free network: the probabilities $P_{out}(k)$ and $P_{in}(k)$ that a document has $k$ outgoing and incoming links follow a power-law over several orders of magnitude, i.e. $P(k)\sim k^{-\gamma}$, with $\gamma_{in}=2.1$ and $\gamma_{out}=2.45$\cite{diam,ibm,small}. Since no complete topological map of the www is available, we limited our study to a subset of the web containing $325,729$ nodes and $1,469,680$ links ($\langle k\rangle=4.59$)\cite{diam}.
Despite the directedness of the links, the response of the system is similar to the undirected networks we investigated earlier: after a slight initial increase, $d$ remains constant in the case of random failures, while it increases for attacks (see Fig.$\,$2c). The network survives as a large cluster under high rates of failure, but the behavior of $\langle s\rangle$ indicates that under attack the system abruptly falls apart at $f_c^w=0.067$ (Fig.$\,$3c).

In summary, we find that scale-free networks display a surprisingly high degree of  tolerance against random failures, a property not shared by their exponential counterparts. This robustness is probably the basis of the error tolerance of many complex systems, ranging from cells\cite{jeong} to distributed communication systems. It also explains why, despite frequent router problems\cite{labovitz}, we rarely experience global network outages or, despite the temporary unavailability of many webpages, our ability to surf and locate information on the web is unaffected. However, the error tolerance comes at the expense of attack survivability: the diameter of these networks increases rapidly and they break into many isolated fragments when the most connected nodes are targeted. Such decreased attack survivability is useful for drug design\cite{jeong}, but it is less encouraging for communication systems, such as the Internet or the www. While the general wisdom is that attacks on networks with distributed resource management are less successful, our results indicate that the topological weaknesses of the current communication networks, rooted in their inhomogeneous connectivity distribution, have serious effects on their attack survivability, that could be exploited by those seeking to damage these systems.

\begin{figure}[htb]
\psfig{figure=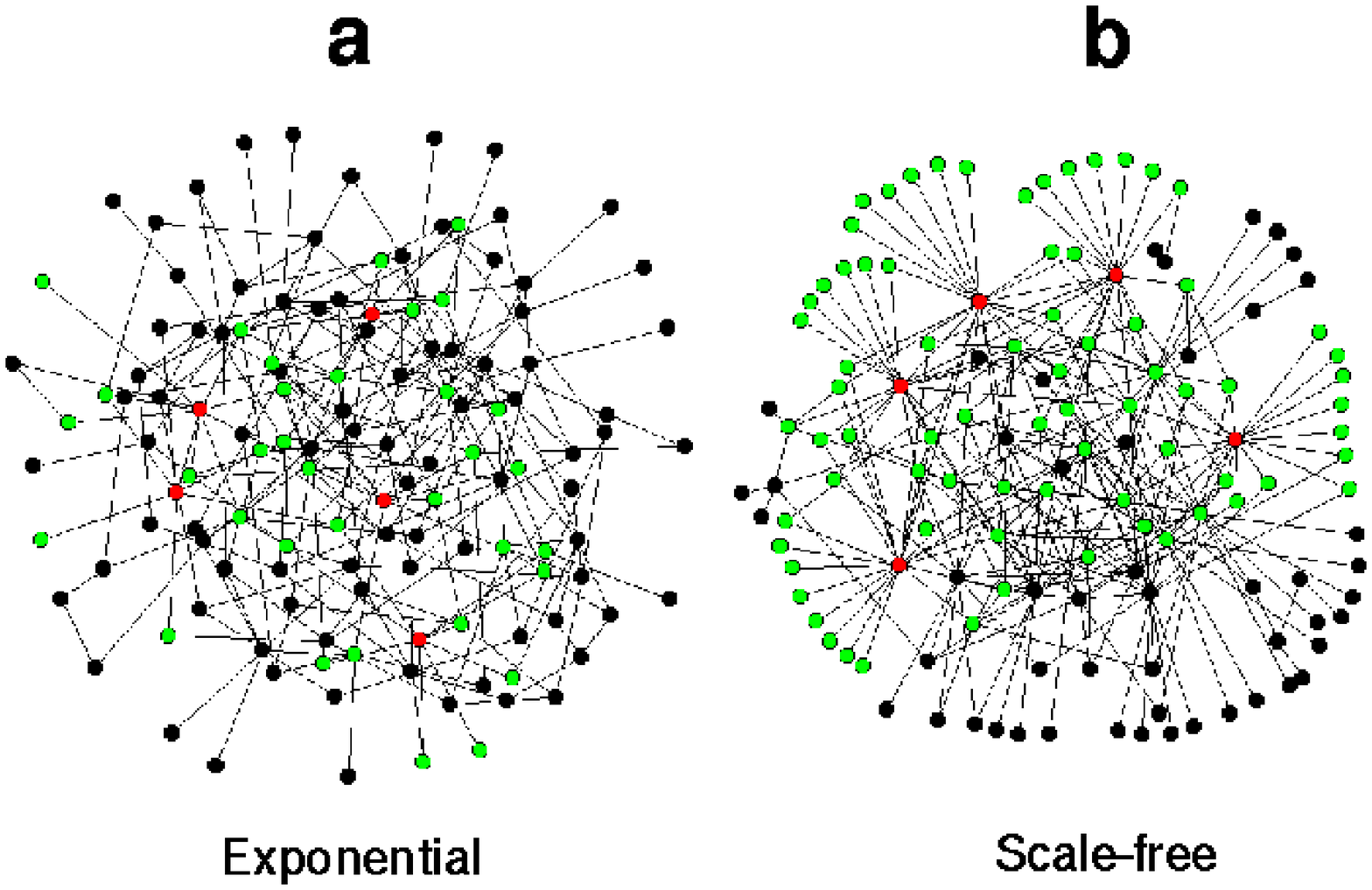,height=4.6in,width=6.9in}
\renewcommand\baselinestretch{0.90}
\caption{ Visual illustration of the difference between an exponential and a scale-free network. The exponential network {\bf a} is rather homogeneous, i.e. most nodes have approximately the same number of links. In contrast, the scale-free network {\bf b} is extremely inhomogeneous: while the majority of the nodes have one or two links, a few nodes have a large number of links, guaranteeing that the system is fully connected. We colored with red the five nodes with the highest number of links, and  with green their first neighbors. While in the exponential network only $27\%$ of the nodes are reached by the five most connected nodes, in the scale-free network more than $60\%$ are, demonstrating the key role the connected nodes play in the scale-free network. Note that both networks contain $130$ nodes and $215$ links ($\langle k\rangle=3.3$). The network visualization was done using the {\it Pajek} program for large network analysis
$<$http://vlado.fmf.uni-lj.si/pub/networks/pajek/pajekman.htm$>$.}
\end{figure}

\newpage

\begin{figure}[h]
\hspace{0.6in}\psfig{figure=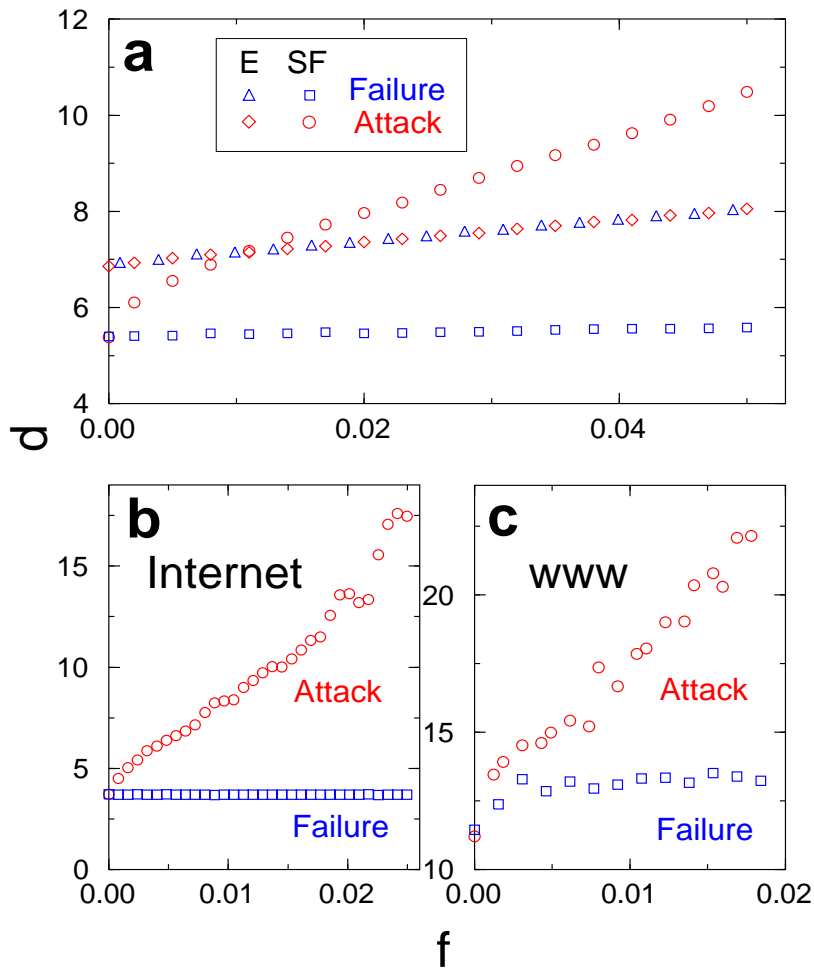,height=6.5in,width=5.1in}
\renewcommand\baselinestretch{0.90}
\vspace{-1in}
\caption{ Changes in the diameter of the network as a function of the fraction of the removed nodes. {\bf a}, Comparison  between the exponential (E) and scale-free (SF) network models, each containing $N=10,000$ nodes and $20,000$ links (i.e. $\langle k\rangle=4$). The blue symbols correspond to the diameter of the exponential (triangles) and the scale-free (squares) network when a fraction $f$ of the nodes are removed randomly (error tolerance). Red symbols show the response of the exponential (diamonds) and the scale-free (circles) networks to attacks, when the most connected nodes are removed. We determined the $f$ dependence of the diameter for different system sizes ($N=1,000$, $5,000$, $20,000$) and found that the obtained curves, apart from a logarithmic size correction, overlap with those shown in {\bf a}, indicating that the results are independent of the size of the system. 
Note that the diameter of the unperturbed ($f=0$) scale-free network is smaller than that of the exponential network, indicating that scale-free networks use more efficiently the links available to them, generating a more interconnected web. {\bf b}, The changes in the diameter of the Internet under random failures (squares) or attacks (circles). We used the topological map of the Internet, containing $6,209$ nodes and $12,200$ links ($\langle k\rangle=3.4$), collected by the National Laboratory for Applied Network Research $<$http://moat.nlanr.net/Routing/rawdata/$>$. {\bf c}, Error (squares) and attack (circles) survivability of the world-wide web, measured on a sample containing $325,729$ nodes and $1,498,353$ links\protect\cite{diam}, such that $\langle k\rangle=4.59$.}
\end{figure}

\newpage

\begin{figure}[htb]
\vspace{-1in}
\psfig{figure=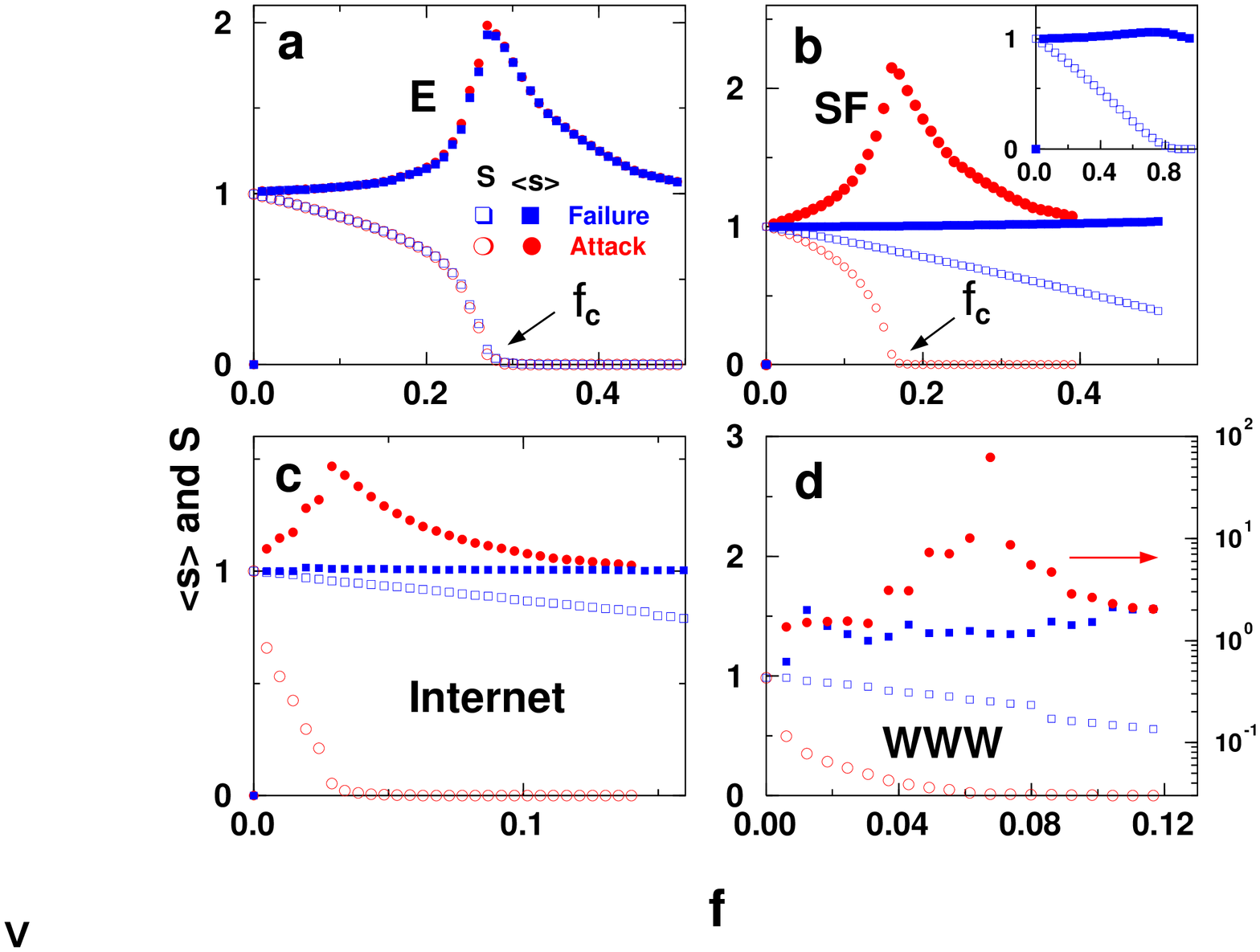,height=5in,width=5.8in,angle=-90}
\renewcommand\baselinestretch{0.90}
\caption{ Network fragmentation under random failures and attacks. The relative size of the largest cluster $S$ (open symbols) and the average size of the isolated clusters $\langle s \rangle$ (filled symbols) in function of the fraction of removed nodes $f$ for the same systems as in Fig.$\,$2. The size $S$ is defined as the fraction of nodes contained in the largest cluster (i.e. $S=1$ for $f=0$). {\bf a}, Fragmentation of the exponential network under random failures (squares) and attacks (circles). {\bf b}, Fragmentation of the scale-free network under random failures (blue squares) and attacks (red circles). The inset shows the error tolerance curves for the whole range of $f$, indicating that the main cluster falls apart only after it has been completely deflated. Note that the behavior of the scale-free network under errors is consistent with an extremely delayed percolation transition: at unrealistically high error rates ($f_{max}\simeq 0.75$) we do observe a very small peak in $\langle s\rangle$ ($\langle s_{max}\rangle\simeq 1.06$) even in the case of random failures, indicating the existence of a critical point. For {\bf a} and {\bf b} we repeated the analysis for systems of sizes $N=1,000$, $5,000$, and $20,000$, finding that the obtained $S$ and $\langle s\rangle$ curves overlap with the one shown here, indicating that the overall clustering scenario and the value of the critical point is independent of the size of the system. Fragmentation of the Internet ({\bf c}) and www ({\bf d}), using the topological data described in Fig.$\,$2. The symbols are the same as in {\bf b}. Note that $\langle s\rangle$ in {\bf d} in the case of attack is shown on a different scale, drawn in the right side of the frame. While  for small $f$ we have $\langle s\rangle\simeq 1.5$, at $f_c^w=0.067$ the average fragment size abruptly increases, peaking at $\langle s_{max}\rangle\simeq 60$, then decays rapidly. For the attack curve in {\bf d} we ordered the nodes in function of the number  of outgoing   links, $k_{out}$. Note that while the three studied networks, the  scale-free model,  the Internet and the www have different  $\gamma$, $\langle k\rangle$ and  clustering coefficient\protect\cite{small_world}, their response to attacks and errors is identical.  Indeed, we find that the difference between these quantities changes  only $f_c$ and the magnitude  of $d$, $S$ and $\langle s\rangle$,  but not the nature of the response of these networks to perturbations.}
\end{figure}

\newpage
\begin{figure}[ht]
\psfig{figure=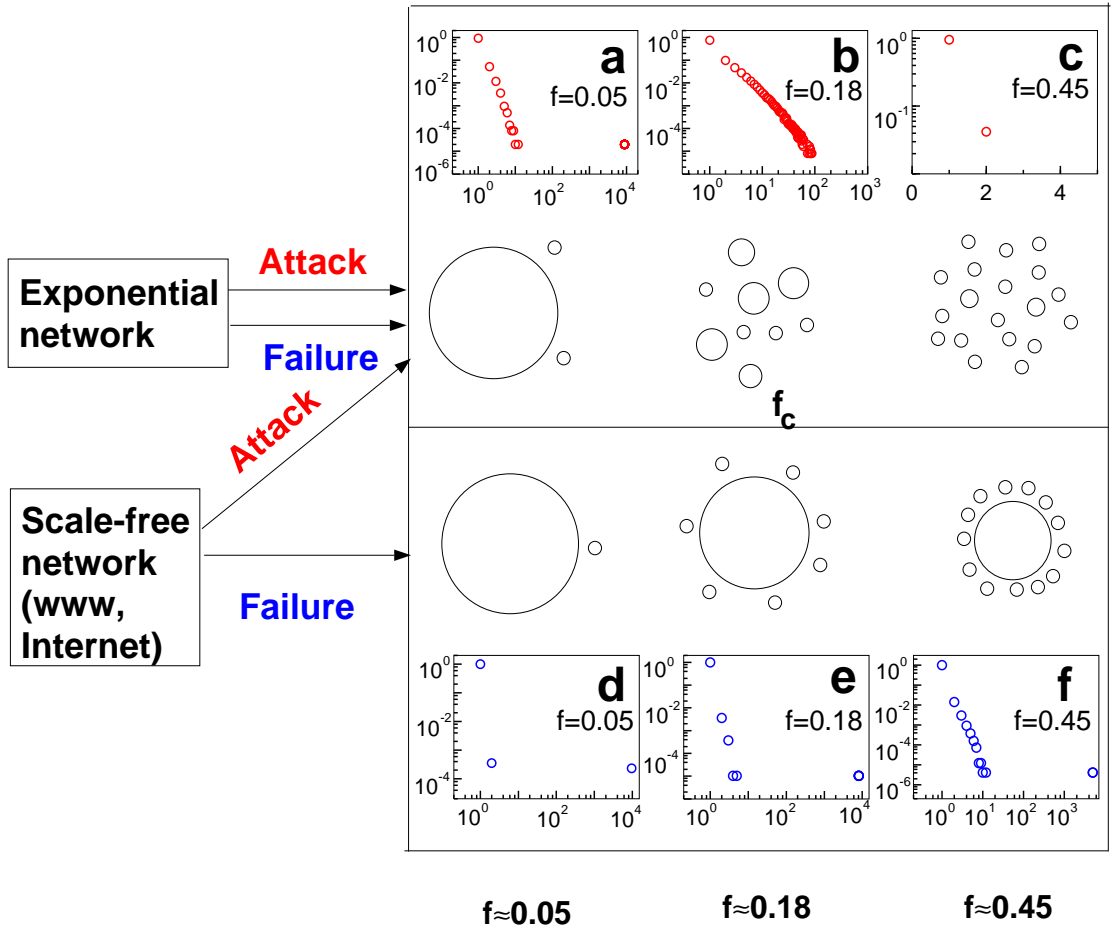,height=8.4in,width=6.3in}
\vspace{-2.5in}
\renewcommand\baselinestretch{0.90}
\caption{Summary of the response of a network to failures or attacks. The insets show the cluster size distribution for various values of $f$ when a scale-free network of parameters given in Fig.$\,$3b is subject to random failures ({\bf a}-{\bf c}) or attacks ({\bf d}-{\bf f}). {\bf Upper panel:} Exponential networks under random failures and attacks and scale-free networks under attacks behave similarly: for small $f$ clusters of different sizes break down, while there is still a large cluster. This is supported by the cluster size distribution: while we see a few fragments of sizes between $1$ and $16$, there is a large cluster of size $9,000$ (the size of the original system being $10,000$). At a critical $f_c$ (see Fig.$\,$3) the network breaks into small fragments between sizes $1$ and $100$ ({\bf b}) and the large cluster disappears. At even higher $f$ ({\bf c}) the clusters are further fragmented into single nodes or clusters of size two. {\bf Lower panel:} Scale-free networks follow a different scenario under random failures: The size of the largest cluster decreases slowly as first single nodes, then small clusters break off. Indeed, at $f=0.05$ only single and double nodes break off ({\bf d}). At $f=0.18$, when under attack the network is fragmented ({\bf b}), under failures the large cluster of size $8,000$ coexists with isolated clusters of size $1$ through $5$ ({\bf e}). Even for unrealistically high error rate of $f=0.45$ the large cluster persists, the size of the broken-off fragments not exceeding $11$ ({\bf f}).}
\end{figure} 

\end{document}